\documentclass[runningheads]{llncs}
\usepackage{graphicx}
\usepackage{amsmath}

\usepackage{caption}

\usepackage{lmodern}  				
\usepackage[utf8]{inputenc}    		
\usepackage[T1]{fontenc} 			
\usepackage[english]{babel}			
\usepackage{xcolor}

\usepackage{amsmath}
\usepackage{amsfonts}
\usepackage{mathtools}

\usepackage[hidelinks]{hyperref}	
\usepackage[a4paper, left=2.5cm, right=2.5cm, top=2.5cm, bottom=2.75cm]{geometry}

\usepackage{enumerate}			
\usepackage{booktabs}

\usepackage{authblk} 			
\usepackage{abstract} 			

\usepackage{wrapfig,lipsum,booktabs}

\usepackage[misc]{ifsym}
\usepackage{fancyhdr}
\usepackage{subfig}

\newcommand\todo[1]{}

\begin{document}
\title{Learning to Prune Instances of Steiner Tree Problem in Graphs}

\author{Jiwei Zhang \and
Deepak Ajwani \orcidID{0000-0001-7269-4150}}
\authorrunning{J. Zhang and D. Ajwani}
%
\institute{University College Dublin, Dublin, Ireland}
\maketitle              
\begin{abstract}
  We consider the Steiner tree problem on graphs where we are given a set of nodes and the goal is to find a tree sub-graph of minimum weight that contains all nodes in the given set, potentially including additional nodes. This is a classical NP-hard combinatorial optimisation problem. In recent years, a machine learning framework called learning-to-prune has been successfully used for solving a diverse range of combinatorial optimisation problems. In this paper, we use this learning framework on the Steiner tree problem and show that even on this problem, the learning-to-prune framework results in computing near-optimal solutions at a fraction of the time required by commercial ILP solvers. Our results underscore the potential of the learning-to-prune framework in solving various combinatorial optimisation problems.

\end{abstract}
\section{Introduction} 
Steiner tree problem is a classical well-studied combinatorial optimisation problem. We consider the variant of this problem on graphs, where we are given an input weighted graph, a set of terminal nodes $V$ and the goal is to compute a minimal connected subgraph that contains all nodes in $V$. This variant of the problem is NP-hard. Furthermore, it is even NP-hard to approximate it within a factor of $96/95$~\cite{inapprox}. Steiner tree problems in graphs are applied to various problems in research and industry~\cite{app0} including network design problems, design of integrated circuits, location problems and more recently even in machine learning, systems biology, and bioinformatics. Given its applications and hardness, numerous approximation algorithms and heuristics have been developed to solve this problem efficiently.

Traditional approaches to solve combinatorial optimisation problems include the usage of integer linear programming solvers, constraint programming approaches, parameterised and approximation algorithms, various heuristics including nature based metaheuristics (e.g., genetic algorithms) and customising algorithms to specific input distribution. In recent years, machine learning techniques have been explored to speed-up the computation of solutions (c.f.,~\cite{survey} for a survey). Machine learning techniques are particularly useful in applications where the same optimisation problem is solved again and again on a regular basis, maintaining the same problem structure but differing in the data \cite{NIPS2017_d9896106}. Many of these learning techniques aim to learn the optimal solution directly. An example of such an end-to-end machine learning technique on Steiner tree problem is the Cherrypick solution by Yan et al.~\cite{heuristic}, where a deep reinforcement learning technique called DQN is used together with an embedding to encode path-related information in order to predict the optimal solution directly. Such end-to-end approaches generally suffer from poor generalisation (resulting in poor solutions for larger and/or more complex problem instances) and poor interpretability and explanability of the algorithms (implicit in the parameters of the deep learning models) learnt. Since in real industry setting, new constraints are routinely added to the problem, poor interpretability means that we do not know if the learnt model will still work with newly discovered constraints and thus, new models have to be learnt from scratch every time this occurs.

  A key reason for the poor generalizability of end-to-end approaches is that they do not leverage any algorithmic insight into the problem, instead relying solely on the input and embedding vectors. This is also an important factor for them needing deep learning models with poor interpretability. To address some of these issues, a learning-to-prune approach~\cite{aaai,clique_arxiv,tsp2,alenex22} has been proposed. Instead of trying to predict the optimal solution directly, it uses a supervised learning model to predict the elements (e.g., nodes/edges) that are unlikely to be part of optimal solution and prune them from further consideration. Once these elements are pruned out, the problem on the remaining elements (predicted to be in optimal solution by the classifier or where the classifier was not confident) is usually quite small and can, thus, be solved using existing exact/approximate approaches. The supervised learning approach leverages a large number of features that can capture the algorithmic insights from the state-of-the-art literature on the problem. As such, classification techniques with significantly fewer parameters, such as random forest, SVM etc. work very well in this framework and there is no need for more complex deep learning models. An added advantage of this framework is that it requires far fewer labelled training instances for training, which is vital for NP-hard optimisation problems.

In this paper, we explore if the learning-to-prune framework can be used to solve the Steiner tree problem efficiently. We show that on the instances of SteinLib dataset where the LP relaxation doesn't return integral solutions, we can obtain optimal or near-optimal solution in orders of magnitude less time compared to solving the ILP formulation directly using Gurobi. 

\section{Applying Learning-to-prune to Steiner tree problem} The key decisions we need to make in order to apply the learning-to-prune framework involve (i) the choice of the ILP formulation for the Steiner tree problem, (ii) the choice of features and (iii) the classification models. We found that the choice of the ILP formulation was crucial for the success of the learning-to-prune framework on this problem. An ILP formulation with smaller integrality gap seems to be particularly suitable for this framework as its LP relaxation can be used as a highly discriminative feature in the process of search-space pruning. Thus, we carefully considered the various ILP formulations for this problem and opted for a formulation based on multicommodity flow transmission~\cite{formulation}.

\subsection{Integer Linear Programming Formulation}
In this formulation, we first convert an undirected Steiner tree problem into a directed version by replacing each edge $\{i,j\}$ with weight $c_{ij}$ by two directed edges $(i,j)$ and $(j,i)$ of the same weight $c_{ij}$. Then, we consider the problem of connecting the set of terminal nodes in the undirected graph as sending a unit flow from an arbitrary terminal node (referred to as root and indexed as node 1) to the remaining terminal nodes in the corresponding directed graph. In particular, the $k^{th}$ flow goes from the root to the $k^{th}$ terminal node (the first flow goes from the root node to itself). Since all flows are moving away from root and towards the terminal nodes, the aggregation of these paths result in the Steiner tree in the undirected graph. Next, we describe this formulation in more detail:

\subsubsection{Sets and Indices}
\begin{itemize}
  \item $i \in N = \{1,2, ..,n \} = \{1\} \cup V \cup S$: The index number of root node is 1. Here, $\{1\} \cup V$ is the set of terminal nodes and $S$ is the set of remaining nodes.
  
  \item $E= \{(i,j)\}$ : Set of directed edges. Note that the size of $E$ is double the size of the edge set of the undirected graph.
 
  \item ${G} = ({N} , {E})$: A graph where the set ${N}$ defines the set of nodes, the set ${E}$ defines the set of directed edges and the set $\{1\} \cup {V} \subseteq N$ defines the set of terminals. 
  
  \item $T \subset E$ : Set of edges that represents a tree spanning $\{1\} \cup V$ in $G$.

  \item $c_{ij} \in {R}^+$: The cost of the arc $(i, j)$, for all $(i, j) \in {E}$.
  
\end{itemize}
\subsubsection{Decision Variables}
\begin{itemize}
  \item $y_{ij} \in \{0, 1\}$: This variable is equal to 1, if edge $(i, j)$ is in the set $T$. Otherwise, the decision variable is equal to zero.
  
  \item $x^k_{ij}$: This is the amount of commodity $k$ (the amount of flow from node $1$ to $k$) that goes through edge $(i, j)$.
\end{itemize}  
\subsubsection{Objective Function}
Minimize the total cost of $T$:

\begin{equation} \label{eq:obj}
minimize \sum_{(i,j) \in E}c_{ij} \cdot y_{ij}
\end{equation}
\subsubsection{Constraints}

\begin{equation} \label{eq:con1}
\sum_{h \in N}x^k_{ih} - \sum_{j \in N}x^k_{ji} = 
\left\{
\begin{aligned}
1 & , & i= 1, \\
-1 & , & i = k, \\
0 & , & i \neq 1, k.
\end{aligned}
\right.
\quad \forall k \in V,
\end{equation}
\begin{equation} \label{eq:con2}
x^k_{ij} \leq y_{ij},
\end{equation}

\begin{equation} \label{eq:con3}
x^k_{ij} \geq 0, \quad \forall (i, j) \in E, k \in V,
\end{equation}

\begin{equation} \label{eq:con4}
y_{ij} \in \{0, 1\}.
\end{equation}

\subsubsection{Constraints Explanation}
As described before, we use an embedded multi-commodity network flow problem to describe the connectivity of the Steiner tree problem. In constraint~\ref{eq:con1}, one unit of commodity k must be transmitted from node 1 to node k. Constraint~\ref{eq:con2} indicates that when the flow is allowed in an edge, the edge must be in the solution. Constraint~\ref{eq:con3} enforces that the flow on any edge is non-negative while Constraint~\ref{eq:con4} enforces that the variables $y_{ij}$ are binary. Constraints~\ref{eq:con1}-~\ref{eq:con4} indicate that a feasible solution must have a directed path of edges (i. e. $y_{ij} = 1$) between node 1 and a node belonging to $V$. 

\subsection{Features}
Another requirement for applying the learning-to-prune framework is to have a set of highly discriminative features that can separate the edges in the optimal solution from those that are not. In other words, we want to associate a set of features with each edge that will allow us to train a classifier separating the set of edges in the optimal solution from the other edges. For the Steiner tree problem, we consider features associated with the LP relaxation, weight of the edges with respect to other edge weights and centrality of associated nodes. Except for Eigenvector centrality, these features can be computed quite fast and they allow us to achieve a high degree of pruning with little loss in objective function value.

\subsubsection{LP relaxation feature} The first feature is the value of edge variables $y_{ij}$ in the LP relaxation of the problem. A high value of this variable suggests a higher likelihood of the edge appearing in the optimal solution. We didn't consider any signals from the dual of the problem and we also didn't utilise the variables $x_{ij}^k$ in this study. Note that the computation of LP relaxation requires significantly less time compared to the ILP computation.

\subsubsection{Weight-Related Features} \label{wgt_feature}
We used the following weight-related features: normalised weight, standardised weight, chi-square of normalised weight and local rank of edge $(i,j)$ at node $i$ and $j$. Here, local rank refers to the rank of this edge in the sorted order of all edges (by weight) incident at the node.

\subsubsection{Centrality Features} \label{cen_feature} To capture the discriminative power of an edge further, we also use the centrality of the incident nodes. Intuitively, the edges between highly central nodes are more likely to be part of optimal Steiner trees as they are crucial for low-weight connectivity. Specifically, we use the following centrality features: degree centrality, betweenness centrality and Eigenvector centrality. For all centrality features, NetworkX is used to construct the graph and calculate the feature values \cite{SciPyProceedings_11}. As these centrality features are focusing on nodes instead of edges, each centrality is calculated as the minimum value and the maximal value of the two incident nodes $i$ and $j$ for each edge $(i,j)$.

Degree centrality is defined as the number of links incident upon a node, which is the simplest centrality feature to calculate. It simply measures the importance of a node by how many edges are incident to it. Betweenness centrality is widely used in weighted graphs as it captures the fraction of shortest paths passing through a given node~\cite{wang2008betweenness}. Eigenvector centrality is also an important centrality feature that captures the ``influence'' of a node in the network: A high eigenvector score means that a node is connected to many nodes who themselves have high scores.

\subsection{Classification} As noted by the previous work on learning-to-prune~\cite{aaai,clique_arxiv,tsp2,alenex22}, the exact classification model is not so crucial in this framework. We experimented with five different classification techniques: \textbf{Random forest} (RF), \textbf{Support vector machine} (SVM), \textbf{Logistic Regression} (LR), \textbf{K-nearest neighbour} and \textbf{Gaussian naive bayes}. While the SVM performs best on this problem, the main insights from the experimental results remain the same for all these classifiers. This provides us confidence that our results are not too specific to a particular classification model, but are more broad-based.

\subsection{ILP on the pruned subgraph}
We run the exact Steiner tree ILP formulation on the unpruned graph so obtained. This can be done by fixing all the edge variables of the pruned edges to 0 in the ILP formulation and solving the modified ILP using an ILP solver. The output of this modified ILP is then returned as the output of the learning-to-prune approach. 

\subsection{Ensuring Feasibility} 
One issue with the learning-to-prune framework is that the pruned set of edges may not contain any feasible solution of the problem that satisfies all constraints. In other words, the pruned graph may have multiple connected components with nodes in $V$ divided between these components. To resolve this issue, we add back all edges for which the corresponding variable has a non-zero value in the LP relaxation. Assuming that the input graph was connected, the set of edges with non-zero values in LP relaxation solution will maintain connectivity among the terminal nodes and thus, with their addition, feasibility will be guaranteed.

Next, we evaluate the quality of this solution as well as the running time of this approach and the relative importance of the different features used in the framework.

\section{Result} 
\subsection{Experimental Setup}
In the benchmark SteinLib~\cite{steinLib} dataset, we found that there are only 55 problem instances for which the LP relaxation of the considered  ILP formulation does not return integral solutions. Thus, we only focus on these instances and select 80\% of them with the smallest running times as training and use the remaining 20\% of the instances with the largest running time as the test dataset. This is because we want to show that our model generalises from smaller instances to larger and more complex instances in this dataset.

To compare the different classification techniques, a 70/30 split is applied and all classifiers are trained on the same training set and tested on the same test set (different from the training set). To have a general comparison between them, the test graphs are all combined into one large dataframe as a general evaluation. This is to test whether the classifier is sensitive to these optimal edges without considering how it performs in a real graph problem.

\subsubsection{Feature Importance}
The feature importance for training the model is shown in Table~\ref{tab:fi}(a) for a SVM classification model and Table~\ref{tab:fi}(b) for a logistic regression classification model\footnote{calculated as the product of the feature coefficient with the standard deviation of feature values in the training set}. Unsuprisingly, Table~\ref{tab:fi}(a) shows that the LP relaxation feature is the most discriminative of all. An important observation here is that even though LP relaxation feature is important, it accounts for less than half of the discriminative power of all the features. This implies that this feature alone isn't enough, but other features also contribute significantly to improving the accuracy of the classification model and the entire learning framework is needed. Similarly, Table~\ref{tab:fi}(b) shows that while the LP relaxation feature is the most discriminative, the other features (such as the maximum degree centrality of the two incident nodes and the minimum local rank of the two incident nodes) also prove to be quite useful in the classification.

\begin{table}[h!]
  \begin{center}
\subfloat[]{\begin{tabular}{|l|r|}
 \hline
Feature &  Importance \\
\hline\hline
LP relaxation feature    &    0.462\\
Normalised Weight          &    0.108\\
Variance                   &    0.083\\
Degree Centrality Max      &    0.052\\
Eigenvector Centrality Max &    0.048\\
Betweenness Centrality Max &    0.048\\
Degree Centrality Min      &    0.046\\
Local Rank j               &    0.041\\
Eigenvector Centrality Min &    0.039\\
Betweenness Centrality Min &    0.038\\
Local Rank i               &    0.036\\
\hline
\end{tabular}}
\quad \hspace*{1cm}
\subfloat[]{\begin{tabular}{|l|r|}
    \hline
Feature &  Importance \\
\hline\hline
LP relaxation feature    &    1.35\\
Normalised Weight          &    0.16\\
Normalised Weight Std                   &    0.12\\
Degree Centrality Max      &    -0.49\\
Eigenvector Centrality Max &    0.11\\
Betweenness Centrality Max &    0.01\\
Degree Centrality Min      &    0.18\\
Local Rank Max               &    -0.92\\
Eigenvector Centrality Min &    0.19\\
Betweenness Centrality Min &    0.02\\
Local Rank Min               &    0.68\\
Edges Connected          & 0.05\\
\hline
\end{tabular}}
\caption{Relative feature importance of different features based on (a) a random forest classification model and (b) a logistic regression classification model \label{tab:fi}}
\end{center}
\end{table}

    

\subsubsection{Comparison of Different Classification Models} 
\begin{figure}[!ht]%
  \subfloat[]{
 \includegraphics[width=0.48\linewidth]{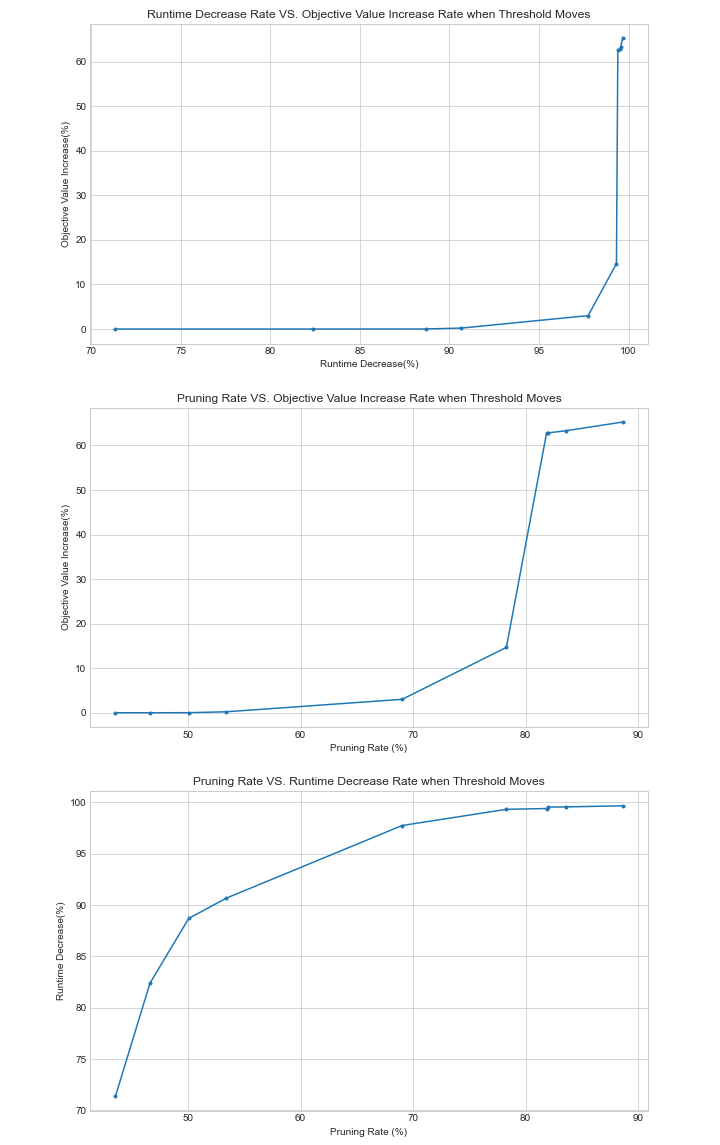}}
  \subfloat[]{
 \includegraphics[width=0.48\linewidth]{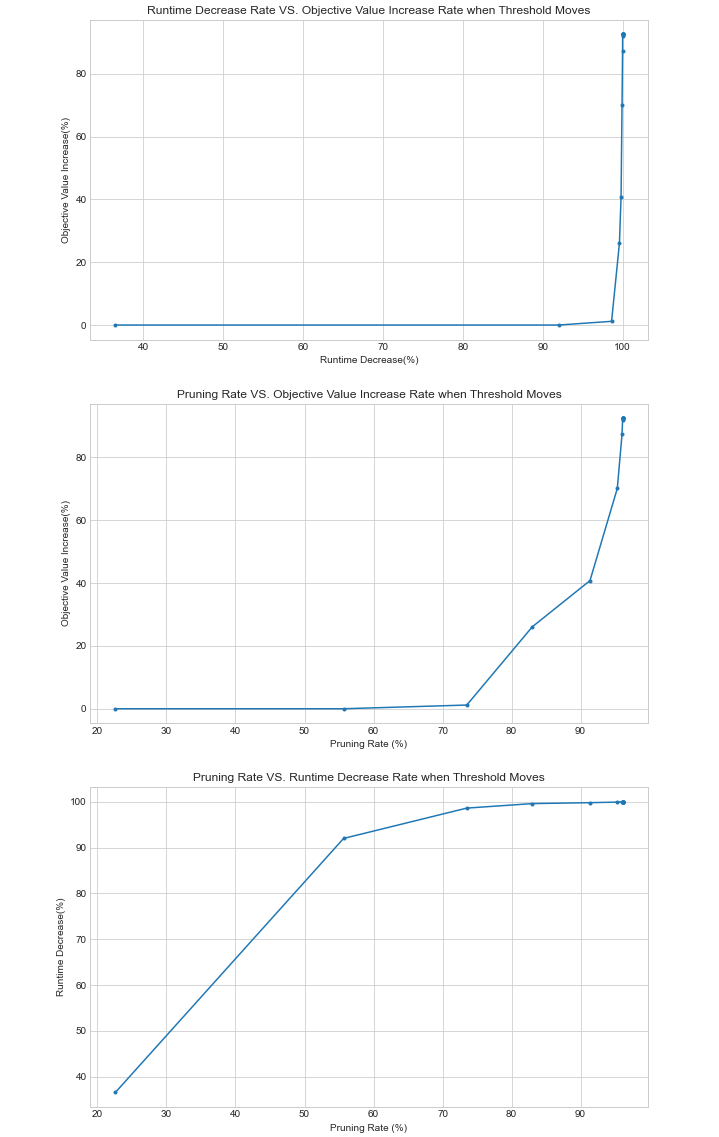}}
 \caption{\label{fig:com} Different trade-offs obtained by varying the threshold of a logistic regression (left) and a SVM (right) classifier}
\end{figure}

To have a general comparison between different classification models for this problem, we test logistic regression and support vector machine by adjusting their thresholds. In this test, we find that both the classifiers are able to decrease the running time drastically with little loss in objective function value. A pruning rate of 60-70\% resulted in a significant reduction in running time while increasing the objective function only slightly. While the general trends are similar across the two classifiers, SVM gives a better trade-off between the objective function value and running time. The comparison is shown in Figure~\ref{fig:com}.

\subsubsection{Objective Function vs. Running Time Trade-off}
As noted in Figure~\ref{fig:com}, both classifiers obtained a drastic reduction in running time at little loss in objective function value. This effect is further illustrated in Table~\ref{tab:Hard}, which presents the results of the learning-to-prune approach on the 10 test instances using the SVM classification model. We first note that on these larger and more complex instances, the time to compute all the features including the LP values is very small compared to the time to run the original ILP. More importantly, the running time of the learning-to-prune approach (including the time to compute features and then running the ILP with hard pruning constraint) is around 99\% less than the original ILP solver time on these instances (using the Gurobi solver). In 7 of these 10 instances, the hard pruning is able to find the optimal solution itself in significantly less time. In the remaining three instances, the resultant increase in the objective function value because of the mistakes in the pruning process is still very small (less than 0.6\%).

\begin{table}[!ht]
    \centering

        \begin{tabular}{l||ccc||cccc}
        
        \hline
        {} &  Objective &  Objective  &  Objective &  Runtime &  Feature Computation &  ILP Solver  &  Runtime \\
        {} & (Original)  &   (After Pruning) &  Increase \% &  (Original) &  Runtime &  Runtime &  Decrease \% \\
        \hline
i160-344   & 8307           & 8324        & 0.20     & 27245.21        & 54.17       & 157.71  & 99.22         \\
i160-244   & 5076           & 5103        & 0.53     & 7762.75         & 25.68       & 47.13   & 99.06          \\
i160-345   & 8327           & 8327        & 0            & 70653.84         & 51.93       & 242.82  & 99.58          \\
i160-343   & 8275           & 8275        & 0            & 20897.36         & 50.54       & 114.90  & 99.21          \\
i160-342   & 8348           & 8355        & 0.08     & 91351.38         & 60.09      & 1384.39 & 98.42          \\
i160-313   & 9159           & 9159        & 0            & 3832.54         & 15.25        & 84.73   & 97.39          \\
i160-241   & 5086           & 5086        & 0            & 6446.48         & 24.78       & 32.78   & 99.11          \\
i160-341   & 8331           & 8331        & 0            & 52473.68          & 53.65       & 104.74  & 99.70           \\
i160-245   & 5084           & 5084        & 0            & 3014.05         & 28.05       & 15.95   & 98.54          \\
i160-242   & 5106           & 5106        & 0            & 4817.80         & 27.34       & 42.81   & 98.54 \\
        \hline
        \end{tabular}

    \caption{Running time and objective function values returned by Gurobi ILP solver and the learning-to-prune framework (with SVM classifier) for different test instances}
    \label{tab:Hard}
\end{table}
\label{detailed_result}

\begin{figure}[!ht]%
  \subfloat[]{
 \includegraphics[width=0.3\linewidth]{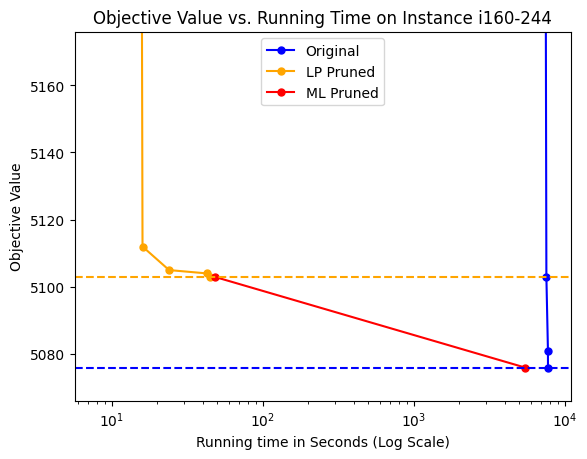}}
  \subfloat[]{
    \includegraphics[width=0.3\linewidth]{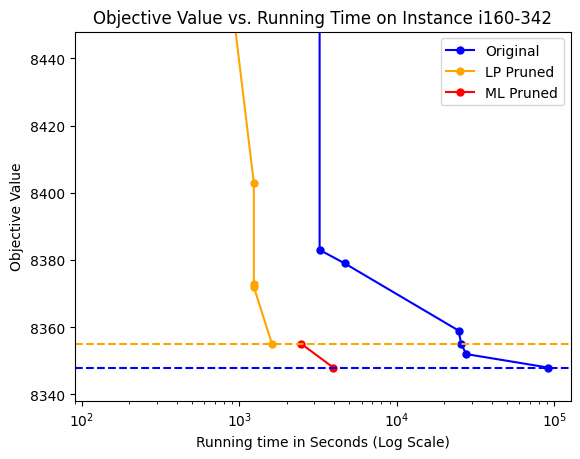}}
  \subfloat[]{
 \includegraphics[width=0.3\linewidth]{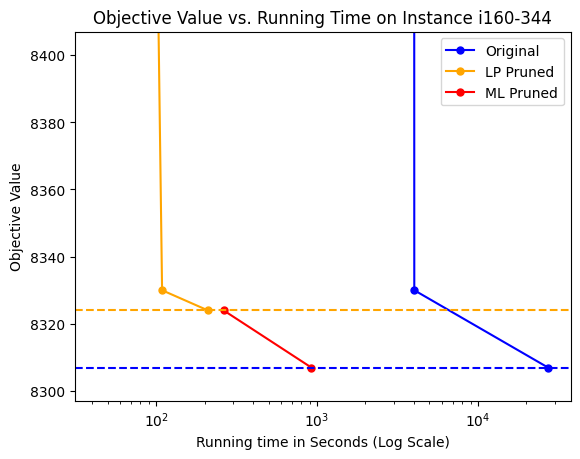}}
 \caption{Comparing hard pruning (referred ``ML Pruned'') with LP-based pruning and Gurobi ILP solver on the original formulation of Steiner tree problem on instances (a) i160-244, (b) i160-342 and (c) i160-344 respectively. The dashed blue horizontal line represents the optimal ILP solution, while the dashed orange horizontal line represents the optimal ILP solution on the instance obtained by pruning all edges with zero LP values. \label{fig:lp-ilp-ml} }
\end{figure}

At this stage, a natural question to ask is how do these results compare with directly pruning based on the LP relaxation values with different thresholds. Next, we consider the three instances where the hard pruning doesn't get optimal results and compare the results of the hard pruning with pruning based on the LP relaxation value. Figure~\ref{fig:lp-ilp-ml} presents the result of such a comparison. We observe that the hard pruning provides significantly better objective value vs running time trade-off compared to the Gurobi ILP solver. In particular, note that Gurobi requires considerably more time to reach a comparable objective function value. In all three instances, hard pruning based on a diverse range of features provides solutions with better objective function values compared to directly pruning based on LP relaxation values, even though it takes some more time. In these plots, the dashed orange horizontal line in these curves represent the objective function value of the optimal ILP solution on the instance obtained by pruning all edges with zero LP relaxation value. Note that even this value is higher than the solutions from our hard pruning approach.

In applications where we wish to reduce the optimality gap even further, we can use the soft pruning approach. The idea here is that instead of adding a hard constraint that no edge can be taken from the set of pruned edges (fixing those edge variables to 0), we add a soft constraint that a small constant number of edges can be taken from the set of pruned edges in the returned solution. In other words, we add the constraint that sum of all edge variables corresponding to pruned edges has to be less than equal to a small constant. This is implemented by simply adding the corresponding constraint in the ILP formulation. While the soft pruning still retains all the edge variables in the ILP formulation, it prunes the search space considerably. When applied on the instance i160-344, the soft pruning approach, that allows just one edge from the pruned set, finds the optimal solution of the original problem. The running time of this approach on this instance is around 3000 seconds, which is still considerably less than the original ILP time of around 27000 seconds, but more than the time of the hard pruning approach (around 150 seconds). 

\section{Conclusion}
Our experiments show that the learning-to-prune framework provides optimal or near-optimal solutions on instances of the SteinLib benchmark at a fraction of the costs of the Gurobi ILP solver. While the feature based on LP relaxation is unsurprisingly the most discriminatory feature for classification, the hard pruning is able to achieve better objective function value compared to pruning directly based on LP relaxation values. It shows that combining the signal from different features using classification models is an effective strategy to prune the problem instances. Our results show the versatility of the learning-to-prune framework in solving various combinatorial optimisation problems.

%
%
%

\end{document}